
%
\input jnl

\preprintno{ILL--(TH)--93--7}
\preprintno{DAMTP--93--15}

\title{Three--Dimensional Quantum Gravity Coupled to Gauge Fields}

\author{Ray L. Renken}

\affil
Department of Physics
University of Central Florida
Orlando, Florida  32816

\author{Simon M. Catterall}

\affil
Department of Applied Math and Theoretical Physics
University of Cambridge
Silver St., Cambridge  CB3 9EW
England

\author{John B. Kogut}

\affil
Loomis Laboratory of Physics
University of Illinois at Urbana--Champaign
1110 West Green Street
Urbana, Illinois  61801

\abstract{ We show how to simulate $U(1)$ gauge fields coupled to
three--dimensional quantum gravity and then examine the phase diagram of this
system.  Quenched mean field theory suggests that a transition separates
confined and deconfined phases (for the gauge matter) in both
the negative curvature phase and the positive curvature phase of the quantum
gravity, but numerical simulations find no evidence for such transitions. }
\vfil
\eject
\doublespace
\pageno=2

\subhead{ Introduction }
In two dimensions, dynamically triangulated random surfaces have proven to
be a useful discretization of quantum gravity.  An analogous approach in
three dimensions has been formulated [1,2].  However, this system can only be
studied in the grand canonical ensemble because no set of ergodic moves have
been found for fixed volume.  The action can be written as
$$S = \alpha N_0 - \beta N_3 \eqno(1)$$
where $N_0$ is the number of nodes and $N_3$ is the the number of tetrahedra
(i.e. the volume).  $\beta$ is therefore the cosmological constant while
$\alpha$ acts like Newton's constant, since at fixed volume varying the
number of nodes varies the curvature.
In two dimensions, the topological classification of manifolds allows one to
prove that at fixed Euler character there is an exponential bound on the number
of triangulations as
a function of the number of triangles.  This implies that a chemical potential
can control the volume.  In three dimensions there is no corresponding
classification and therefore no corresponding proof of an exponential bound.
The naive result is factorial growth which would mean no partition function
could be defined.  It was a significant discovery that a chemical potential
($\beta$) does control the volume in three dimensions [3,4].  The volume
diverges at a particular critical value of $\beta$.  Given that the
thermodynamic limit can be taken, the next issue is the nature of the phase
diagram.  In particular, is there a second order phase transition where a
non--trivial continuum limit
could be taken?  It turns out that there are two phases separated by a
transition at $\alpha^*$.  For smaller values of $\alpha$ the system is in a
negative curvature phase where the coordination number of the nodes diverges.
For larger values of alpha, the system is in a positive curvature phase where
the number of sites is proportional to the volume.  Unfortunately, the
transition between these two phases, where one might have hoped for a continuum
limit with zero curvature, appears to be first order so that there is no
interesting continuum limit.

This doesn't necessarily spell doom, since an expanded phase diagram (with more
coupling constants) might still contain an appropriate transition.
A simple and interesting
way to expand the phase diagram is to couple matter to the system.  Recently,
studies of Ising matter coupled to three--dimensional quantum gravity have
been completed [5-7].  If the matter is placed on the nodes, the phase diagram
consists of the gravitational transition and an Ising transition.  The two
cross at zero Ising coupling so that the Ising matter does not change the
nature of the gravitational transition.  If a dual formulation
is employed in which the matter fields are placed on the
elementary volumes, the gravitational transition is also unchanged.

In this paper, we couple gauge fields to three--dimensional quantum gravity.
(Our simplicial lattices correspond to triangulations of $S^3$).
Aside from the possibility of finding a non--trivial continuum limit, this
system is interesting because it couples gauge fields and quantum gravity.
In two--dimensions, the analogous system is topological --- quantum gravity
has no interesting effect [8].  Three dimensions is thus the lowest
dimension in which coupling these two theories gives a non--trivial result.
In what follows, we restrict ourselves to the U(1) gauge theory.

\subhead{ Mean Field Theory }
A quenched mean
field theory proved useful in the case of Ising matter coupled to
three--dimensional quantum gravity and it proves useful for gauge matter as
well.
The action for U(1) gauge fields coupled to three--dimensional quantum gravity
is
$$S = \alpha N_0 - \beta N_3 + \lambda \sum_{\Delta} \cos \phi_{\Delta}
\eqno(2) $$
where the sum is over all triangles and $\phi$ is the sum of gauge angles
about the given triangle.  A simple approximation to the phase diagram can
be obtained by doing mean field theory for the gauge variables while fixing
the volume and the number of nodes.  This allows us to consider the influence
of the background metric on the matter, but neglects the influence of the
matter on the metric.  The simplest mean field theory [9] for gauge theories
begins with a trial action
$$S_H = H\sum_{\rm links} \cos\theta\eqno(3)$$
for which the free energy per link is simple
$$u(H) = \ln I_0(H)\eqno(4)$$
where $I_0$ is a modified Bessel function.
Adding and subtracting the trial action to the gauge part of the full action
gives
$$Z = {\rm Tr}e^S = Z(H)<\exp(\lambda\sum_{\Delta}\cos\phi_{\Delta} -
 H\sum_{\rm links}\cos\theta)>_H$$
$$ \ge Z(H) \exp<\lambda\sum_{\Delta}\cos\phi_{\Delta} -
H\sum_{\rm links}\cos\theta>_H\eqno(5)$$
where $Z$ is the partition function for the gauge part of the full action,
$Z(H)$ is the partition function for the trial action and the subscript $H$
denotes that the trial action is used as the weight in the expectation values.
The resulting free energy per unit volume is
$$F(H,\lambda) = (u N_1 - H u^{\prime} N_1 + \lambda (u^{\prime})^3 N_2) / N_3
\eqno(6)$$
(the prime denotes the derivative with respect to $H$, $N_1$ is the number of
links, and $N_2$ is the number
of triangles).
This must be maximized with respect to the field $H$.
Maximizing gives a first order transition at $\lambda^* = 0.794$ in the
negative
curvature phase and $\lambda^* = (1 + 1/c)*0.794$ in the positive curvature
phase, where $c = N_3/N_0$ as the volume is taken large.  (A more sophisticated
gauge invariant approach leads to the same results). This result suggests the
thick line in
the phase diagram in figure 1.  Note that mean field theory cannot be taken
too seriously and is given here just to guide the later numerical calculations.
In three dimensions on a fixed lattice, for instance, the theory can be
rewritten as a Coulomb gas that is in an ionized phase for all temperatures.
Mean field theory mistakes a crossover for a phase transition.

Placement of the gravitational transition in the phase diagram is guided by
contact with the pure gravity theory that can be made in the limits of large
and small $\lambda$.  Note that (using $N_1 = N_0 + N_3$)
$$\prod_{\rm links} \left( \int_0^{2\pi} d\theta \right) = (2\pi)^{N_1}
= e^{N_1\ln 2\pi} = e^{(N_0 + N_3)\ln 2\pi} \eqno(7)$$
so that
$$Z(\alpha,\beta,\lambda = 0) =
  Z_{\rm pure~grav}(\alpha + \ln 2\pi,\beta - \ln 2\pi)\eqno(8)$$
implying that the gravitational transition is shifted to
$\alpha^* \sim 1.91 - 2.16$ [4].
For very large $\lambda$ the theory also approaches a pure gravity theory.
In this limit, (since $N_2 = 2 N_3$)
$$\sum_{\Delta} \lambda \cos (\phi_{\Delta}) \sim \lambda N_2 = 2
\lambda N_3\eqno(9)$$
implying
$$Z(\alpha,\beta,\lambda) \sim Z_{\rm pure~grav}(\alpha,\beta - 2\lambda)
\quad {(\lambda {\rm~such~that} \cos\phi_{\Delta} \sim 1)} \eqno(10)$$
This final equation suggests that the gravitational transition is only
weakly dependent on $\lambda$ as indicated by the dotted line in figure 1.

\subhead{ Detailed Balance }
The task now is to numerically check figure 1 and to determine the order of
the phase transitions suggested by mean field theory.  In order to simulate
this theory,
we must derive the detailed balance relations in the grand--canonical ensemble.
In three dimensions, any triangulation can be reached from any other
triangulation through a series of local moves chosen from a set of four
possibilities.  These moves are labelled by the number of tetrahedra before
and after the move.  The (1,4) move (figure 2) corresponds to
the insertion of a node into a tetrahedron
and the (4,1) move is its inverse.  The (2,3) move (figure 3)
replaces a triangle
separating two tetrahedra with a new link at the boundary of three tetrahedra.
The (3,2) move is its inverse.
In contrast to the situation with Ising matter, where only the (1,4) detailed
balance relations were modified by the addition of matter, here all the
relations are modified.  First, consider the (2,3) move.  Here, it
is necessary to assign a value for the gauge field on a
potential new link.
Label the state with two tetrahedra $A$ and the state with the
new link $\theta$, where $\theta$ is an angle between $0$ and $2\pi$.  Then
consider attempting a $2\rightarrow 3$ move with 50\% probability and a
$3\rightarrow 2$ move with 50\% probability.  In the $2\rightarrow 3$ case
choose the triangle by picking a tetrahedron at random and then choosing
one of its four sides.  Note that there are two ways of picking the same
triangle with this procedure. In the $3\rightarrow 2$ case, the link is
chosen from a list of all possible links with coordination number 3.  If
$l_3$ is the number of such links, the resulting detailed balance equation is
$${1\over 2} {1\over N_3} {1\over 4} 2 e^{S_A} P(A\rightarrow \theta) =
 {1\over 2} {1\over l_3} e^{S_{\theta}} P(\theta\rightarrow A) \eqno(11)$$
The acceptance rate can be increased by integrating over the possible
values of $\theta$ so that it is first decided whether to update the
triangulation or not and then later the gauge field values are determined.
Integrating over all possible values of $\theta$ and assuming that
$P(\theta \rightarrow A)$ can be chosen independent of $\theta$ gives
$${l_3 \over 2 N_3} P(A\rightarrow B) = P(B\rightarrow A) \int_0^{2\pi}
e^{S_{\theta} - S_A} d\theta \eqno(12)$$
where B denotes the state with a new link regardless of its value.  Define
$$I_{23} = \int_0^{2\pi} e^{S_{\theta} - S_A} d\theta  \eqno(13)$$
(which will turn out to be a gauge invariant function of the loops on the
surface of the cluster of the two or three tetrahedra under consideration) and
denote the ratio
$$R_{23} = l_3 / 2 N_3 \eqno(14)$$
then detailed balance is satisfied by
$$P(A\rightarrow B) = { I_{23} \over R_{23} + I_{23} } \eqno(15)$$
$$P(B\rightarrow A) = { R_{23} \over R_{23} + I_{23} } \eqno(16)$$

Now consider detailed balance for the (1,4) move.  In this case, attempt
a $1\rightarrow 4$ move or a $4\rightarrow 1$ move each with a 50\%
probability.  In the former case, just pick a tetrahedron at random.  In
the latter case, pick from a list of sites known to have coordination
number four.  Detailed balance requires
$${1\over 2} {1\over N_3} e^{S_A} P(A\rightarrow \Theta) =
 {1\over 2} {1\over l_4} e^{S_{\Theta}} P(\Theta\rightarrow A) \eqno(17)$$
where $A$ represents the state with no node, $\Theta$ represents the four
gauge degrees of freedom living on the four new links, and $l_4$ represents
the number of nodes with coordination number four.  Again, define
$$I_{14} = \int e^{S_{\Theta} - S_A}\eqno(18)$$
where now integration is over four angles, and
$$R_{14} = l_4 / N_3\eqno(19)$$
then detailed balance is satisfied by
$$P(A\rightarrow B) = { I_{14} \over R_{14} + I_{14} } \eqno(20)$$
$$P(B\rightarrow A) = { R_{14} \over R_{14} + I_{14} } \eqno(21)$$
where now $B$ represents the presence of a node without specifying the four
new gauge degrees of freedom associated with it.  These update probabilities
tell us whether to put in or take out a link and whether to put in or take
out a node.  They do not indicate what the gauge degrees of freedom should be.
This is taken care of by a subsequent Metropolis or heat bath step.  One
caveat should be added.  The ratio in $R_{14}$ is
interpreted differently depending on whether a $1\rightarrow 4$ move is being
contemplated or a $4\rightarrow 1$ move is being contemplated.  In the former
case, $N_3$ is the number of tetrahedra to begin with while $l_4$ is the number
of sites with coordination number $4$ if the move is accepted.  In the latter
case, $l_4$ is the number of sites with coordination number $4$ to begin
with while $N_3$ is the number of tetrahedra there will be if the move is
accepted (i.e. the current number of tetrahedra minus $3$). Similar
considerations are required in the case of the (2,3) move.

At first sight, these equations suggest that a simulation of gauge fields
coupled to quantum gravity in this grand--canonical ensemble would be
impractical.  An integration needs to be done every time there is a local
update.  We will see that the integral $I_{23}$ is not really very troublesome.
The integral $I_{14}$ is much more demanding computationally, but it turns
out to be within the realm of practicality.

Consider first the action difference for
the former integral (the (2,3) move).
$$S_{\theta} - S_A = -\beta + \lambda ( t_1 + t_2 + t_3 - t_0) \eqno(22)$$
where 1, 2, and 3 label the new triangles associated with the new link and
0 labels the old triangle that has been eliminated. The
quantities $t_i$ are the cosines
of the sum of the gauge angles about the specified loop.  The needed gauge
integral is therefore
$$I_{23} = e^{-\beta -\lambda t_0} \int_0^{2\pi}
e^{ \lambda (t_1 + t_2 + t_3) } d\theta\eqno(23)$$
If the links, $\alpha_i$, involved in the integral are labeled as in figure 3
and $x_i$ is defined such that
$$x_1 \equiv \alpha_1 + \alpha_2, \quad
x_2 \equiv \alpha_3 + \alpha_4, \quad
x_3 \equiv \alpha_5 + \alpha_6\eqno(24)$$
then
$$t_1 + t_2 + t_3 = \cos(\theta + x_1) + \cos(\theta + x_2) +
\cos(\theta + x_3) \eqno(25)$$
$$ = r \cos(\theta + \omega)\eqno(26)$$
where $\omega$ is independent of $\theta$ and
$$r^2 = \left( \sum_{i=1}^3 \cos x_i \right)^2 + \left( \sum_{i=1}^3
\sin x_i \right)^2 \eqno(27)$$
Finally,
$$I_{23} = 2\pi\ I_0(\lambda r)\ e^{-\beta -\lambda t_0} \eqno(28)$$
Is this result gauge invariant?  The values of the gauge fields enter only
through $r$.  It is easy to see that $r$ is gauge invariant.  Change variables
to
$$z_1 = x_1 - x_2, \quad z_2 = x_1 - x_3, \quad z_3 = x_2 - x_3 \eqno(29)$$
then
$$r^2 = 3 + 2 (\cos z_1 + \cos z_2 + \cos z_3 ) \eqno(30)$$
Since each of the $z_i$ is a gauge loop, $r$ is manifestly gauge invariant.

$I_{14}$ is more complicated.  Here there are four gauge variables to be
integrated over, labeled as in figure 2.  The action difference is
$$S_{\Theta} - S_A = \alpha - 3 \beta + \lambda \sum_{i=1}^{i=6} \cos p_i
\eqno(31)$$
where the $p_i$ are the new plaquettes.  In terms of the gauge variables
$$p_1 = \theta_1 + \theta_2 + x_1\eqno(32)$$
$$p_2 = \theta_1 + \theta_3 - x_5\eqno(33)$$
$$p_3 = \theta_1 - \theta_4 - x_3\eqno(34)$$
$$p_4 = \theta_2 - \theta_3 - x_4\eqno(35)$$
$$p_5 = \theta_2 + \theta_4 - x_2\eqno(36)$$
$$p_6 = \theta_3 + \theta_4 - x_6\eqno(37)$$
The action ratio can now be rewritten
$$e^{S_{\Theta} - S_A} = e^{\alpha - 3 \beta}\prod_{i=1}^{i=6}\left(
\sum_{k_i = -\infty}^{\infty} I_{k_i}(\lambda ) e^{ik_i p_i}\right)\eqno(38)$$
The exponential can be rewritten as $i$ times
$$\theta_1(k_1 + k_2 + k_3) + \theta_2(k_1 + k_4 + k_5) +
\theta_3(k_2 - k_4 + k_6) + \theta_4(-k_3 + k_5 + k_6) $$
$$+ x_1k_1 - x_5k_2 - x_3k_3 - x_4k_4 - x_2k_5 - x_6k_6\eqno(39)$$
Integration over $\theta_1$, $\theta_2$, and $\theta_3$ gives a factor
$$(2\pi)^3 \ \delta(k_1 + k_2 + k_3) \ \delta(k_1 + k_4 + k_5)
\ \delta(k_2 - k_4 + k_6)\eqno(40)$$
Notice that these three $\delta$--functions automatically make the coefficient
of $\theta_4$ zero.  This is a consequence of gauge invariance under
transformations at the center node.  Integration over $\theta_4$ just gives
another $2\pi$.  The $\delta$--functions give
$$k_3 = - k_1 - k_2, \qquad k_5 = - k_1 - k_4, \qquad k_6 = k_4 - k_2
\eqno(41)$$
so that the $x_i$ terms become
$$k_1(x_1 + x_2 + x_3) + k_2(- x_5 + x_3 + x_6) + k_4(- x_4 + x_2 - x_6)
\eqno(42)$$
Each coefficient of the $k_i$ is a gauge invariant loop on the surface of
the tetrahedron.  If the coefficients of the $k_i$ are relabeled as $z_1$,
$z_2$, and $z_3$ respectively then the final result is
$$I_{14} = (2\pi)^4 e^{\alpha - 3\beta} \sum_{m_1=-\infty}^{\infty}
\sum_{m_2=-\infty}^{\infty} \sum_{m_3=-\infty}^{\infty} I_{m_1}(\lambda) $$
$$I_{m_2}(\lambda) I_{m_3}(\lambda) I_{m_1+m_2}(\lambda) I_{m_1+m_3}(\lambda)
I_{m_3-m_2}(\lambda) e^{i( m_1z_1 + m_2z_2 + m_3z_3) } \eqno(43)$$
Invariance under sign changes in $m_1$, $m_2$, and $m_3$ make it clear that
the right hand side is real --- the exponent can be replaced with a cosine.
(Computations are faster if the expression is rewritten to take advantage
of this fact).

\subhead { Results }
One parameter that has to be tuned is the ratio, $r$, of update sweeps of the
gauge field to the update sweeps of the metric.  In two dimensions, when bosons
are coupled
to quantum gravity, the expectation value of the bosonic action can be
calculated exactly due to scale invariance.  This can be used to check whether
the bosons are in equilibrium and then the equilibrium of the gravity sector
can be checked by varying the above ratio and checking to see that the results
are stable.  Here, there is no analytical check of either the matter or the
gravity sector.  We can only check various expectation values as a function of
the ratio and look for stability.  For a fixed lattice (that has not been
warmed up) one update pass on the gauge fields is insufficient.  We suspect
that on a dynamical lattice, where the connectivity is constantly changing,
that more than one gauge update is required per update sweep through the
metric.  Consider the expectation values of the plaquette, the specific
heat, and the number of nodes, as well as the value of $\beta$ required to
maintain a given volume.  As $r$ is varied from 1 to 16, the values of all of
these quantities change, but reach a plateau near $r = 8$.
This is illustrated in fig. 4 with a plot of the expectation value
of the number of nodes versus $r$.  The expectation value of the plaquette
has a similar behavior, varying from 0.62 to 0.67 as r varies from 1 to 16,
but varying by less than a part per thousand as $r$ is increased from 8 to
16.  $r = 8$ is the value we used.
A number of runs at $r = 1$ were also done.  The results differed
numerically from the $r = 8$ results, but they implied the same conclusions
described below.

The phase diagram suggested by the quenched mean field theory and the partition
function relations can be tested numerically.  The easiest thing to
check is the implication of eqn. (10) that the gravitational transition is
independent of $\lambda$.  As in the case of pure gravity [4], the expectation
value of the number of nodes scales with the volume as
$$<N_0>\sim N_3^{\delta}\eqno(44)$$
where $\delta$ is one in the phase with finite coordination number and less
than one in the phase where the coordination number diverges.  For
$\lambda = 2$, we computed $\delta$ for $\alpha = 1.6$, $\alpha = 1.8$,
and $\alpha = 2.1$ on lattices of volume 2000, 4000, 8000, and 16000.  The
results for $\delta$ are 0.55(1), 0.84(2), and
1.007(1) respectively.  The error bars are statistical only.  The confidence
level of the (log) fits to a straight line were not high, suggesting that
larger lattices are necessary to reach scaling, but the trend is consistent
with a gravitational transition still in the region $\alpha = 1.91 - 2.16$
of the $\lambda = 0$ transition.

The gauge transition is more difficult.  To search for the gauge transitions
we computed the gauge part of the specific heat as a function of $\lambda$ for
various fixed values of $\alpha$ on both sides of the gravitational transition
and looked for a peak growing with volume.  Figures 5 and 6 show the results.
There is a maximum in the specific heat in the region predicted by mean field
theory, but there is no sharp peak, it does not grow with volume, and there
is no evidence for critical slowing down.  The same
results were obtained for large and smaller values of $\alpha$.  This
suggests that the gauge sector is in a strong coupling phase for all $\lambda$.
Other operators show consistent behavior.   For instance, the expectation
value of the plaquette is almost a straight line as a function of $\lambda$ in
the region of $\lambda$ we considered and had no significant volume dependence
on lattices up to a volume of 16000.
One could attempt to verify that the entire phase is in strong coupling
by examining Wilson loops for large $\lambda$
and testing for an area law, but this is much more difficult on a
dynamical triangulation than on a fixed hypercubic lattice and we have not
tried to do it.

A simpler way to try to get information about what is going on in the gauge
sector is to look at monopoles and their clusters.  Consider a
three--dimensional elementary volume (a tetrahedron here) with a
surface of oriented plaquettes, $P_i$ (i.e. $\sum P_i = 0$).  For each $P_i$
define $n_i$ such that
$$P_i = 2\pi n_i + \phi_i \eqno(45)$$
where $-\pi < \phi < \pi$.  Then the monopole number in this volume is
$$m = \sum_i n_i \eqno(46)$$
Generally, a certain value of $n_i$ requires $p_i > (2 n_i - 1)\pi$.  Summing
this inequality over the $s$ sides of the elementary volume gives a limit
on the monopole number:
$$m < s/2 \eqno(47)$$
For a cube, this gives the familiar result that $m \in \{ -2,-1,0,1,2 \}$.
For a tetrahedron there are fewer possibilities: $m \in \{ -1,0,1 \} $.
A monopole density is defined as
$$\rho = (N_+ + N_-) / V \eqno(48)$$
where $N_m$ is the number of tetrahedra with monopole number $m$ and $V$ is
the volume (the number of tetrahedra).  Clusters are defined as groups of
neighboring monopoles (signs are ignored and monopoles are neighbors if the
surfaces of their tetrahedra share a face).  An interesting order parameter for
clusters is the ratio
$$M = \left< {N_{\rm max} \over N_{\rm tot}} \right> \eqno(49)$$
where $N_{\rm max}$ is the number of monopoles in the largest cluster and
$N_{\rm tot}$ is the total number of monopoles.  $M$ is a good order parameter
for a percolation transition, a transition from a phase with many clusters to
a phase with one cluster.  In the three-dimensional quantum gravity plus $U(1)$
gauge fields system, the monopole density is $1/3$ when
$\lambda = 0$ and it slowly and smoothly decreases as $\lambda$ is increased.
(On a cubic lattice the monopole density is 0.43 at $\lambda = 0$ and decreases
for larger $\lambda$.)
The density has very little sensitivity to the value of $\alpha$.  The
cluster order parameter, $M$, is small for $\lambda = 0$
and $\alpha$ in the region of the gravitational transition and it decreases
smoothly as $\lambda$ is increased.  Eventually it increases again, but this
is only because the density  becomes so small that there are of order unity
monopoles (i.e. $N_{\rm max} \sim N_{\rm tot}$).  $M$ also has no sensitivity
to the value of $\alpha$ and this is surprising.  In general, one expects a
low percolation threshold for lattices with a high coordination number
(such as those generated in the negative curvature phase).  Apparently, it
is the coordination number of the tetrahedron that counts (since that is
where the monopole lives) rather than that of the sites.  A tetrahedron has a
coordination number of four, the same
as a site in a two--dimensional square lattice, so one might guess that the
threshold for the tetrahedral lattice is roughly that of the square lattice,
namely 0.59 [10].  This guess can be confirmed by artificially placing
monopoles of a given density on the dynamical lattices and measuring the
threshold directly.  Consequently, the physical monopole density is always
below the percolation threshold and they provide no useful signal.

In conclusion, we have shown how to couple gauge fields to three-dimensional
quantum gravity and made an effort to find an interesting continuum limit.
No appropriate phase transition was found in the space of actions we
considered.

This work was supported, in part, by NSF grant PHY 92--00148.  Some
calculations
were performed on the Florida State University Cray Y--MP and at the Pittsburgh
Supercomputer Center.  We thank Olle Heinonen for discussions.
\references

[1]  N. Godfrey and M. Gross, \prd 43, R1749, 1991.

[2]  M. E. Agishtein and A. A. Migdal, \journal Mod. Phys. Lett., A6, 1863,
1991 and PUPT--1272.

[3]  J. Ambjorn and S. Varsted, NBI--HE--91--17.

[4]  D. V. Boulatov and A. Krzywicki, LPTHE Orsay 91/35.

[5]  R. Renken, S. Catterall, and J. Kogut, \np B389, 277, 1992.

[6]  J. Ambjorn and S. Varsted, NBI--HE--91--45,6.

[7]  C. F. Baillie, COLO-HEP-279.

[8]  J. Wheater, \pl B223, 451, 1989 and \pl B264, 161, 1991.

[9]  C. Itzykson and J.-M. Drouffe, {\it Statistical Field Theory}, Cambridge
University Press, 1989.

[10] D. Stauffer and A. Aharony, {\it Introduction to Percolation Theory},
Taylor \& Francis, 1992.

\endreferences

\figurecaptions

[1] The quenched mean field diagram for three--dimensional quantum gravity
coupled to U(1) gauge fields.

[2]  The tetrahedra involved in a (1,4) move.

[3]  The tetrahedra involved in a (2,3) move.

[4]  The expectation value of the number of nodes, $N0$, versus the ratio of
the number of gauge update sweeps to the number of gravity update sweeps.
$\alpha = 2.1$, $\lambda = 1.0$, and $N3 = 4000$.

[5]  The gauge component of the specific heat as a function of $\lambda$ at
$\alpha = 1.8$ for two lattice sizes: $N_3 = 2000$ (crosses) and $N_3 = 4000$
(boxes).

[6]  The gauge component of the specific heat as a function of $\lambda$ at
$\alpha = 2.2$ for two lattice sizes: $N_3 = 2000$ (crosses) and $N_3 = 4000$
(boxes).
\endfigurecaptions

\endit